\documentclass[aps,epsfig,floats,pre,twocolumn,showpacs]{revtex4-1}
\usepackage{graphicx}
\usepackage[latin2]{inputenc}
\usepackage{amsmath}
\usepackage{amssymb}
\usepackage{epsfig}
\usepackage{bm}
\usepackage{color}

\begin{document}

\title{Enhanced diffusion and anomalous transport of 
magnetic colloids driven above a two-state flashing potential}

\author{Pietro Tierno$^{1,2}$}
\email{ptiernos@ub.edu}
\author{M. Reza Shaebani$^3$}
\email{shaebani@lusi.uni-sb.de}
\affiliation{$^{1}$Departament d'Estructura i Constituents de la Mat\`eria, 
Universitat de Barcelona, 08028 Barcelona\\
$^{1}$Institut de Nanoci\`encia i Nanotecnologia, IN$^2\!$UB, 
Universitat de Barcelona, Barcelona, Spain\\
$^{1}$Department of Theoretical Physics, Saarland 
University, D-66041 Saarbr\"ucken, Germany}

\begin{abstract}
We combine experiment and theory to investigate 
the diffusive and subdiffusive dynamics of paramagnetic colloids driven 
above a two-state flashing potential. The magnetic potential was realized 
by periodically modulating the stray field of a magnetic bubble lattice 
in a uniaxial ferrite garnet film. At large amplitudes $H_0$ of the 
driving field, the dynamics of particles resembles an ordinary random 
walk with a frequency-dependent diffusion coefficient. However, 
subdiffusive and oscillatory dynamics at short time scales is observed 
when decreasing $H_0$. We present a persistent random walk model to 
elucidate the underlying mechanism of motion, and perform numerical 
simulations to demonstrate that the anomalous motion originates from 
the dynamic disorder in the structure of the magnetic lattice, 
induced by slightly irregular shape of bubbles.
\end{abstract}

\maketitle

\section{Introduction\label{sec:introduction}}
Transport and diffusion of microscopic particles through periodic potentials
is a rich field of research from both fundamental and technological points of
view \cite{Reim02,Han09}. Investigation of the particle motion along ordered 
\cite{Reim02} or disordered \cite{San04,Han12} energy landscapes helps to
better understand the dynamics in more complex situations, such as Abrikosov 
\cite{Vil03,Von05} and Josephson vortices in superconductors \cite{Maj03,Ust04}, 
cell migration \cite{Mah09}, or transport of molecular motors \cite{Jul97}. 
Moreover, a periodic potential can be used to perform precise particle sorting
and fractionation \cite{Kor02,Mac03,Hua04,Lac05,Tie08}, thus, being of significant
impact in diverse fields in analytical science and engineering which make use
of microfluidic devices. Colloidal systems provide an ideal opportunity to
investigate different transport scenarios, because of having particle sizes
in the visible wavelength range and dynamical time scales which are experimentally
accessible. In order to force colloidal particles to move along periodic
or random trajectories, static potentials can be readily realized by using
optical \cite{Dal11}, magnetic \cite{Tie07}, or electric fields \cite{Per08}. 
Dynamic landscapes (obtained by periodically or randomly modulating the potential)
are a subject of growing interest since a rich dynamics can be induced due to
the presence of competing time scales. Moreover, flashing potentials where
static landscapes are modulated in time, are usually employed to study
molecular systems \cite{Ast94,Pro94,Cao04} or as an efficient way to transport
and fractionate Brownian species \cite{Fau95,Lib06}.

\begin{figure*}[t]
\begin{center}
\includegraphics[width=\textwidth]{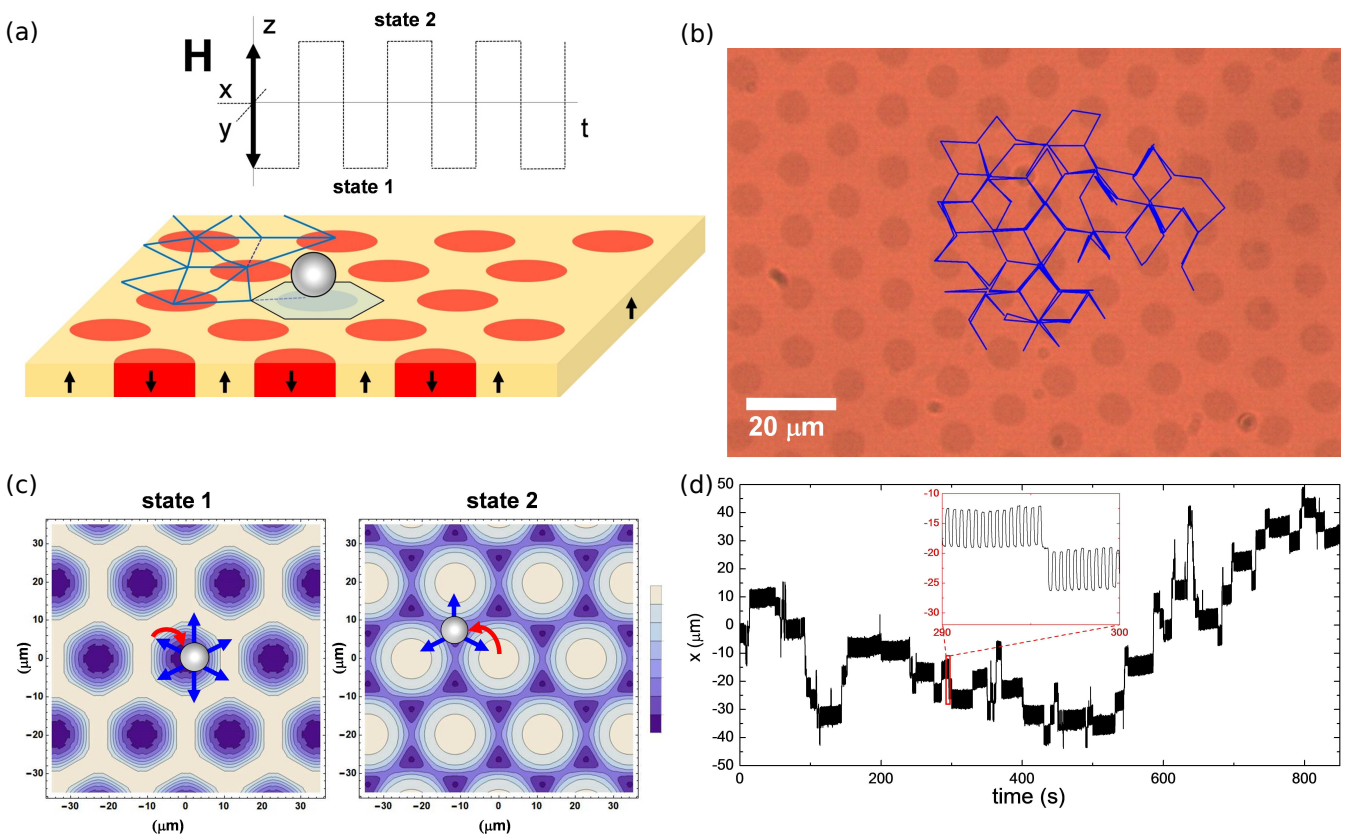}
\caption{(a) Schematic showing a paramagnetic particle driven above a magnetic
bubble lattice by a square wave magnetic field $\bf H$. One Wigner-Seitz cell is
shaded in blue. (b) Microscope image of a bubble lattice with a superimposed
trajectory of one particle (blue lines) for $\omega{=}12.6\,{\rm rad\,s^{-1}}$
and $H_0{=}0.14 \, {\rm M_s}$. (c) Contour plots of the normalized magnetostatic 
energy $E/k_BT$ of one particle above the bubble lattice at two different times 
separated by half a period. Energy maxima (minima) are colored in white (blue). 
The blue arrows indicate the possible paths the particle can undertake in the 
next jump. The red arrows show the path undertaken by the particle to reach the
energy minimum. (d) Time evolution of the $x-$coordinate of one paramagnetic 
particle subjected to an external square wave field with $H_0 {=} 0.14 \, 
{\rm M_s}$ and $\omega {=} 15.7 \, {\rm rad \, s^{-1}}$. The red inset zooms 
in one piece of the trajectory.}
\label{Fig:1}
\end{center}
\end{figure*}

Here we present a combined experimental and theoretical study focused on the
dynamics of microscale particles driven above a flashing magnetic potential.
This potential is generated by periodically modulating the energy landscape
created by an array of magnetic bubbles. As it was previously reported \cite{Tie09}, 
this experimental system is able to generate different dynamical
states depending on the applied field parameters, such as frequency or 
amplitude of the external field. In particular we report the observation 
of enhanced diffusive dynamics at high field strength $H_0$, while the 
motion at lower $H_0$ is subdiffusive with a crossover to normal diffusion 
at long times. At low values of $H_0$, the lattice structure is slightly
disordered during the switching of the magnetic field direction. The resulting
randomness is dynamic, i.e.\ not reproducible after a cycle of external 
drive, and enhances with decreasing $H_0$. By means of numerical simulations 
we verify that in the presence of lattice disorder, the particles frequently 
experience oscillatory motion in local traps. The back and forth motion of 
particle in local traps happens more frequently as the dynamic disorder in 
the structure of the magnetic lattice increases. The trapping events change 
the statistics of the turning angles of the particle from an isotropic 
distribution (limited to the directions allowed by the lattice structure) 
to an anisotropic one with a tendency towards backward directions. Using a 
persistent random walk model, we show that anomalous diffusion arises when 
the turning-angle distribution of a random walker is asymmetric along the 
arrival direction. When the walker has the tendency towards backward directions,
the resulting antipersistent motion is subdiffusive or even strongly 
oscillatory at short time scales. However, the walker has a finite range 
memory of the successive step orientations, i.e.\ the direction gets 
randomized after long times and the asymptotic behavior is ordinary 
diffusion with a smaller long-term diffusion coefficient compared to 
an ordinary random walk. We obtain good agreement between the
analytical predictions, simulations, and experiments.

\begin{figure*}
\begin{center}
\includegraphics[width=\textwidth]{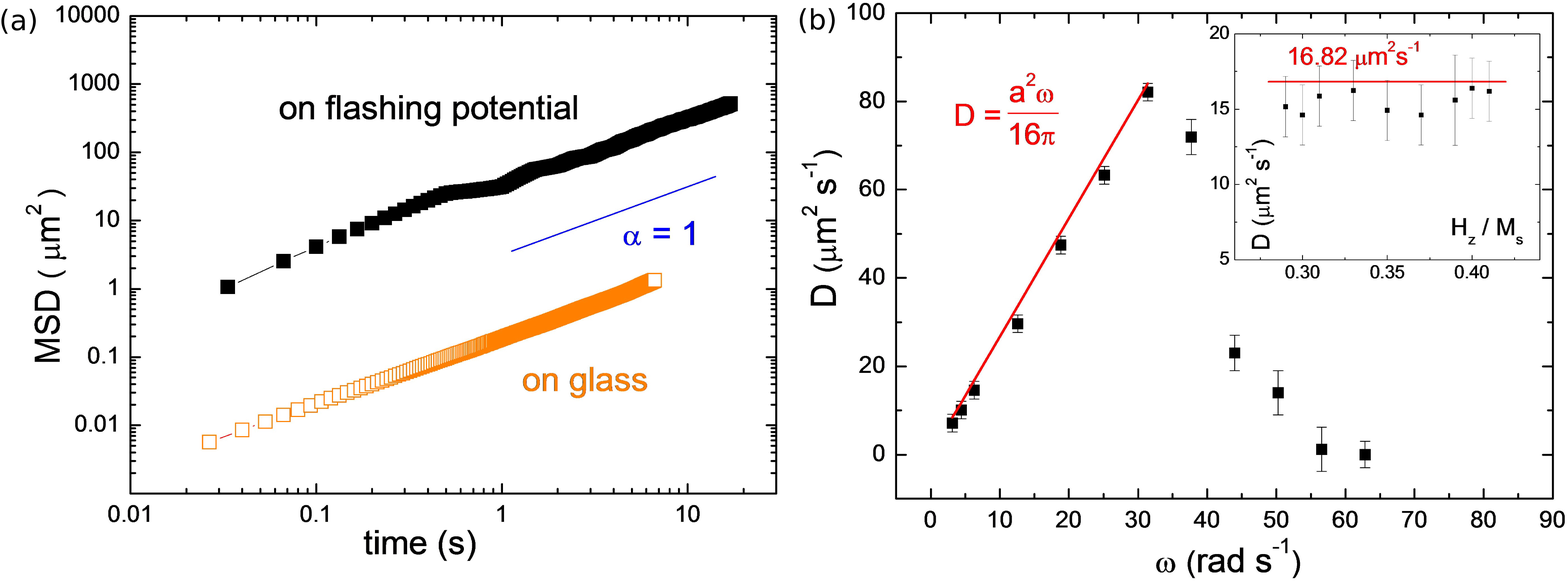}
\caption{(a)
Mean squared displacement versus time for a paramagnetic colloid above a glass 
substrate (open squares) and above a two-state flashing potential (solid squares), 
under a field with amplitude $H_0 {=} 0.37 \, {\rm M_s}$ and angular frequency 
$\omega {=} 6.3 \, \, {\rm rad \, s^{-1}}$. (b) Effective diffusion coefficient 
$D$ vs angular frequency $\omega$ for a colloidal particle driven by an applied 
field with $H_0 {=} 0.37 \, {\rm M_s}$. The solid line denotes the relationship 
$D {=} \frac{a^2\omega}{16 \pi}$. The inset shows $D$ vs $H_0$, at a constant 
angular frequency of $\omega {=} 6.3 \, {\rm rad \, s^{-1}}$.}
\label{Fig:2}
\end{center}
\end{figure*}

The paper is organized in the following manner: First we introduce the setup
in Sec.~\ref{Sec:Setup}. Section \ref{Sec:ExpResults} contains the experimental
results obtained at different field parameters. In Sec.~\ref{Sec:Simulation},
the results of numerical simulations for transport in the presence of dynamic
disorder are discussed and compared with the corresponding experimental data.
The motion of particles is modeled at the level of individual steps via an
antipersistent random walk approach in Sec.~\ref{Sec:PRW}, and finally
Sec.~\ref{Sec:Conclusions} concludes the paper.

\section{Experimental Setup}
\label{Sec:Setup}
A schematic illustrating the experimental system is shown in Fig.~\ref{Fig:1}(a). 
The colloidal particles used are polystyrene paramagnetic microspheres (Dynabeads 
M-270, Invitrogen) having diameter $d{=}2.8 \, {\rm  \mu m}$ and magnetic volume 
susceptibility $\chi {\sim} 0.4$. The particles are diluted in high-deionized 
water and let sediment above the periodic potential generated by a 
bismuth-substituted ferrite garnet film (FGF). The FGF has composition 
Y$_{2.5}$Bi$_{0.5}$Fe$_{5-q}$Ga$_q$O$_{12}$ ($q{\in}[0.5,1]$) and was previously 
grown by dipping liquid phase epithaxy on a $0.5 mm$ thick gadolinium gallium 
garnet substrate \cite{Tie09}. The film has thickness $\sim {\rm 4\mu m}$ and
saturation magnetization $M_s{=}1.7 {\times} 10^4 \, {\rm A m^{-1}}$. In the 
absence of external field, this FGF is characterized by a labyrinth of stripe 
domains with alternating magnetization and a spatial periodicity of $\lambda=9.8 
\,\mu m$. This pattern is converted into a periodic lattice of cylindrical magnetic 
domains by using high frequency magnetic field pulses applied perpendicular to 
the film, with amplitude $H_0$ and oscillating at angular frequency $\omega$. 
As shown in Fig.~\ref{Fig:1}, the cylindrical domains, also known as ``magnetic 
bubbles" \cite{Esc80}, are ferromagnetic domains with radius $R{=}4.2\, {\rm 
\mu m}$, having the same magnetization direction and arranged into a triangular 
lattice with lattice constant $a{=}11.6\, {\rm \mu m}$. We can visualize both 
the magnetic domains in the film and the particles using polarization 
microscopy, due to the polar Faraday effect.

The external oscillating field is obtained by connecting a coil perpendicular 
to the film plane ($z$-direction) with a wave generator (TT{\it i} - TGA1244) 
feeding a power amplifier (IMG STA-800). The custom-made coil was mounted on 
the stage of a polarization light microscope (Nikon, E400) equipped with a 
$100\times$, $1.3$NA objective and a $0.45$ TV lens. The movements of the 
particles are recorded at $60$ fps for $\sim 30$ min with a $\text{b}{/}\text{w}$ 
CCD camera inside an observation area of $179 {\times} 231 \, {\rm \mu m^2}$. 
We then use particle tracking routines \cite{Cro96} to extract the particle 
trajectories and later calculate correlation functions. 

\section{Particle dynamics in a flashing potential}
\label{Sec:ExpResults}
Once deposited above the garnet film, the paramagnetic particles pin to the 
Bloch walls, which are located at the boundary of the magnetic bubbles. To 
generate a two-state flashing potential we apply to the FGF film a square-wave 
modulation of type
\begin{equation}
{\bf H}=H_0 \, \mathrm{sgn} \big(\sin{(\omega t)}\big) {\bf e}_z \, ,
\end{equation}
where $H_0$ is the field amplitude, $\omega$ the angular frequency, and 
$\mathrm{sgn}(x)$ denotes the sign function. The applied field periodically 
changes the radii of the bubbles in the FGF, increasing (decreasing) the size 
of the bubbles when it is parallel (resp.\ antiparallel) to their magnetization 
direction, thus alternating between two distinct states. One can understand 
the effect of the applied field on the energy landscape by calculating the 
magnetostatic potential at the particle elevation \cite{Sob08} see 
Fig.~\ref{Fig:1}(c). In particular, when the field expands the bubbles ({\it 
state 1}), the energy displays a paraboloid-like minimum within the magnetic 
domains. Thus, the magnetic colloids are attracted towards the center of the 
bubble domains. When the field has opposite polarity, it shrinks the size of 
the bubbles and enlarges the interstitial region ({\it state 2}). In this
situation, the magnetostatic potential features six regions of energy minima 
with triangular shape at the vertices of the Wigner-Seitz cell around each 
bubble.

As a consequence, during the transition {\it 1}${\rightarrow}${\it 2}, a 
particle can jump to $6$ possible places [Fig.~\ref{Fig:1}(c), left], while 
in the transition {\it 2}${\rightarrow}${\it 1} the possibilities reduce 
to $3$ [Fig.~\ref{Fig:1}(c), right]. Since we apply a field perpendicular 
to the sample plane (no tilt), the potential preserves its spatial symmetry, 
i.e.\ there is no net drift motion as it was induced in Ref.~[17] by using 
a precessing field. We analyze the particle dynamics by measuring the mean 
squared displacement via a temporal moving average $\text{MSD}(t) {=} \langle 
\big(x(t{+}t') {-} x(t')\big)^2 \rangle {\sim} t^{\alpha}$. Here, $x$ denotes 
the position of the particle projected along one of the crystallographic axes 
and $\alpha$ the exponent of the power-law which is used to distinguish the 
diffusive ($\alpha{=}1$) from subdiffusive ($\alpha{<}1$) dynamics. In our 
system we observe both types of dynamics, which are discussed in the 
following subsections.

\subsection{Diffusive dynamics}
In Fig.~\ref{Fig:2}(a) we compare the MSD for a paramagnetic colloid freely 
diffusing on a glass plane in the absence of FGF film and the one which 
is strongly driven by the flashing potential. Both cases exhibit a normal 
diffusion but with different diffusion coefficients. From the experimental
data of the MSD, the effective diffusion coefficient of the particle can 
be estimated as $D {=} \lim\limits_{t\rightarrow\infty} \text{MSD}(t)/2dt$, 
with $d$ being the spatial dimension (Here $d{=}1$ since the data is 
projected along one of the crystallographic axes). We find that $D$ in 
the presence of the flashing potential ($D {=} 14.6\,{\rm \mu m^2 \, 
s^{-1}}$) is enhanced by nearly two orders of magnitude with respect to 
the one on a glass plate ($D {=} 0.14 \, {\rm \mu m^2 \, s^{-1}}$). In 
this regime of motion, which is observed for field amplitudes $H_0\in 
[0.28, 0.42] \, {\rm M_s}$ and frequency $\omega {<} 35 \, {\rm rad \, 
s^{-1}}$, the particle moves from one domain to the next without performing 
continuous oscillation around a single site, thus, it performs an ordinary 
random walk on a triangular lattice. The step length $l$ of the 
walker can be estimated to be given by one side of the Wigner-Seitz 
cell, i.e.\ $\ell{=}\frac{a}{2} {=} 5.8 \, {\rm \mu m}$, and the 
component of the MSD along a crystallographic axis equals $\langle 
x^2 \rangle{=}(\frac{\ell}{\sqrt{2}})^2{=}\frac{\ell^2}{2}$. Each step 
takes a half-period of the magnetic field pulse, thus, the duration 
of each step is given by $t_s {=} \frac{T}{2} {=} \frac{\pi}{\omega}$.
One thus obtain the diffusion coefficient as \cite{Mac83,Kla00} $D {=}
\frac{\langle x^2 \rangle}{2t_s} {=} \frac{l^2}{4t_s} {=} \frac{a^2
\omega}{16 \pi}$. We confirm this relation in Fig.~\ref{Fig:2}(b) by
measuring $D$ versus the frequency and amplitude of the applied field. 
While $D$ increases linearly with frequency, it decrease rapidly beyond 
$\omega {\sim} 30 \, \text{rad} \, s^{-1}$ since the overdamped particle 
is unable to follow the fast vibrations of the potential, and reduces 
significantly its diffusive dynamics. In contrast to frequency, $D$ 
is almost independent of the amplitude of the applied field, since 
the lattice constant of the magnetic bubble array does not change 
significantly for amplitude $H_0 {<} 0.5 \, {\rm M_s}$ \cite{Tie07}. 
Beyond this value, however, the magnetic bubble lattice starts melting 
and transport of the particles is not possible anymore.

\subsection{Subdiffusive dynamics}
\begin{figure}[t]
\begin{center}
\includegraphics[width=\columnwidth,keepaspectratio]{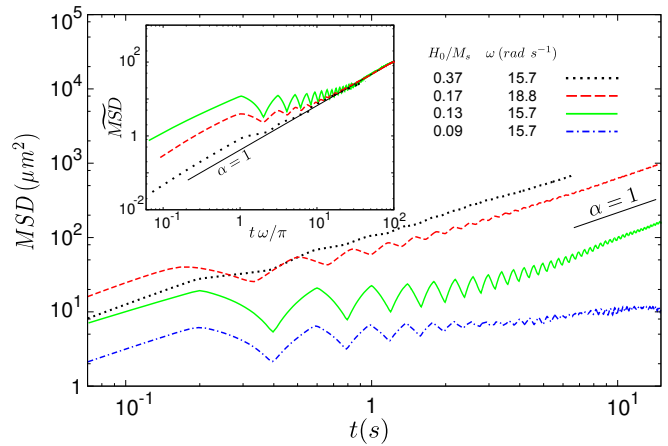}
\caption{Mean squared displacement versus time for a paramagnetic colloid
subjected to a two-state flashing potential for different values of amplitude
$H_0$ and frequency $\omega$ of the applied magnetic field. Inset: The dimensionless
mean squared displacement $\widetilde{\text{MSD}}=\text{MSD}/(\frac{4\ell}{v}
D_{\text{asymp}})$ versus the scaled time $t\omega/\pi$.}
\label{Fig:3}
\end{center}
\end{figure}

Our experimental setup allows us to independently vary both the amplitude
$H_0$ and the frequency $\omega$ of the driving field. At strong fields 
($H_0{>}0.28 \, {\rm M_s}$), the particle jumps between nearest bubbles 
with an equal probability to choose any of the possible movements, thus, 
performs a normal random walk on a triangular lattice ($\alpha{\simeq}1$ 
on all time scales). In contrast, when the field is weak ($H_0 {<} 0.09 
\, {\rm M_s}$), the landscape deformations induced by the applied field 
are small and the particle is unable to leave the magnetic bubble. The 
corresponding particle trajectory is composed of simple oscillations 
between the center of one bubble and one of its six surrounding energy 
minima. In such a pure localization the MSD saturates rapidly, leading 
to an exponent $\alpha{\simeq}0$. In the intermediate regime of amplitudes, 
i.e.\ $H_0 {\in} [0.09,0.28] \, {\rm M_s}$, we observe a subdiffusive 
dynamics at short time scales with a crossover to normal diffusion 
at long times. Figure~\ref{Fig:3} shows several examples of the
temporal evolution of MSD for different values of $H_0$ and $\omega$. 
The initial anomalous exponent varies between $0$ and $1$ depending 
on the strength $H_0$ of the applied magnetic field. While in a previous
work \cite{Tie09} it was found a stable subdiffusive exponent $\alpha{=}
0.5$ for $H_0=0.13 \, {\rm M_s}$ and at different frequencies, the 
value of $\alpha$ may vary by changing the amplitude of the switching 
field. To see this more clearly, we rescale both axes in such a way 
that the asymptotic diffusive regimes of the curves collapse on a 
master curve (see inset). This can be achieved if the MSD is scaled 
by $\frac{4\ell}{v}D_{\text{asymp}}$ ($D_{\text{asymp}}$ being the 
long-term diffusion coefficient, $\ell$ the step size, and $v$ the 
velocity of the particle), and the time axis is scaled by 
$\frac{\pi}{\omega}$ to synchronize the oscillations. Only the 
samples which reach the long-term diffusion limit within our 
experimental time window are considered. With increasing $H_0$, the 
curves are initially more steep and converge faster to the asymptotic 
limit.

Indeed, the lattice structure is slightly disordered at weak fields,
because of the non-uniform deformations of the magnetic bubbles during 
their periodic expansion/contraction. These deformations arise from the 
presence of pinning sites in the film and other inhomogeneities (such as 
dislocations or magnetic and non-magnetic inclusions present in the FGF 
crystal). These defects exert an influence on the Bloch walls analogous 
to the action of a frictional force against the motion of the walls \cite{Bar77}. 
Furthermore, when the field is applied, the magnetic bubbles interact 
via long range dipolar forces \cite{Ses92} and small deviations from 
the 2D projected circular shape induce a slight distortion on the 
triangular lattice. For the fields used in our experiments, these
distortions are not strong enough to create permanent defects in the 
film (such as disclinations or dislocations), but slightly vary the
lattice spacing from place to place. Thus, the spatial structure of
the lattice is not perfectly symmetric in practice. In the presence
of disorder, the particle may oscillate around each site before it 
moves to the next domain (see Fig.~\ref{Fig:1}(d) for an example of
such movements), which leads to subdiffusive dynamics at short time 
scales. However, this bias decays with time and the directions of the 
particle motion become asymptotically randomized, leading to diffusive
dynamics at long times. With decreasing $H_0$, the lattice disorder
and thus the frequency of trapping events enhances, which further
decreases the initial anomalous exponent from one (ordinary random 
walk) towards zero (pure localization).

The frequency $\omega$ of the driving field does not affect the 
MDS behavior; it only rescales the time because the time step is given 
by $t_s {=} \frac{\pi}{\omega}$. Thus, it can be understood why the 
crossover time to asymptotic diffusion was found to follow a power-law 
$\tau_c {\sim} \omega^{-1}$ with $\omega$ \cite{Tie09}.

\section{Simulation results}
\label{Sec:Simulation}
In the previous section we explained the origin of the disorder in the 
lattice structure during switching of the magnetic field. As a result,
the particle spends part of the time in local traps, where it experiences
an oscillatory movement. In a back and forth motion, the particle chooses 
nearly backward directions when it turns. Therefore, the turning-angle 
distribution $f(\phi)$ changes from an isotropic one for ordinary random 
walk on the lattice (limited to the allowed directions by the lattice 
structure) to an anisotropic one with a tendency towards backward 
directions with respect to the current direction of motion. In the
extreme limit of pure localization (remaining permanently in a trap), 
$f(\phi)$ is a delta function at $\phi {=} \pi$. Such anisotropic 
turning-angle distributions slow down the spread of colloids on the 
lattice and cause subdiffusion at short times. The structural 
irregularities in the magnetic bubble lattice are more pronounced 
at smaller values of $H_0$. Hence, with decreasing $H_0$ the chance 
of trapping events increases and the resulting $f(\phi)$ becomes more 
anisotropic, which decreases the initial anomalous exponent.

\begin{figure}[t]
\begin{center}
\includegraphics[width=\columnwidth,keepaspectratio]{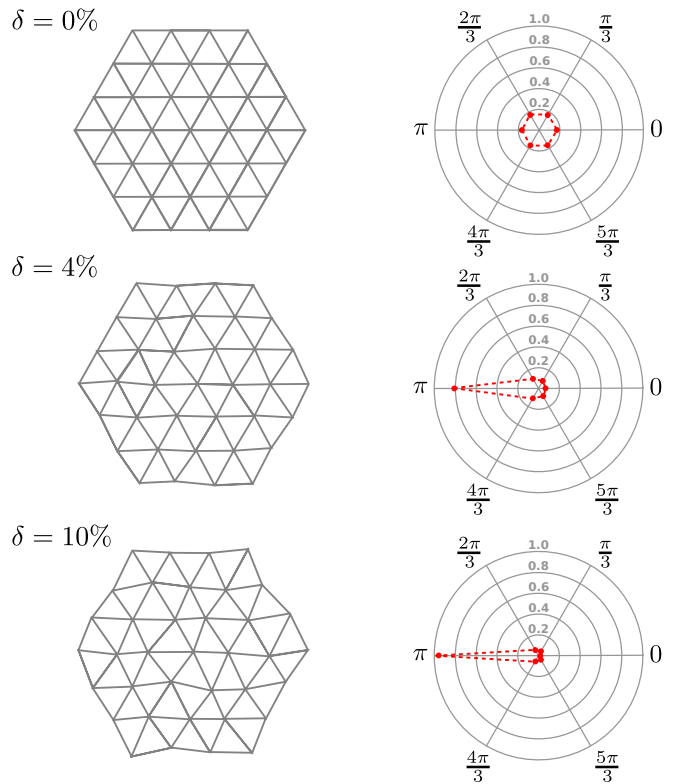}
\caption{(left) Schematic drawing of the disordered lattice
and (right) a polar representation of the corresponding turning-angle
distribution $f(\phi)$ after $10000$ steps in simulations for different 
values of the lattice disorder $\delta$.}
\label{Fig:4}
\end{center}
\end{figure}

In order to ensure that the lattice disorder causes asymmetric 
turning-angle distribution $f(\phi)$ and subdiffusive dynamics, 
we perform Molecular Dynamics simulations of motion on a 
dynamic disordered triangular lattice of synchronous flashing 
magnetic poles. To obtain smooth particle trajectories and a 
detailed time evolution of the MSD, in particular to monitor 
the oscillatory dynamics at short times, MD simulations are 
advantageous compared to other possible methods such as Monte 
Carlo simulations. The simulation cell consists of nearly 
3000 magnetic poles and a colloidal particle which is initially 
released near the center of the system to avoid boundary 
effects. We impose periodic boundary conditions, and consider 
in-plane magnetic interactions between the immobile poles and 
the magnetic colloid. The time step of our in-house code is 
chosen to be $\Delta\,t{=}1{\times}10^{-4}$s, so that the time 
$t_s{=}\frac{\pi}{\omega}$ between two successive switching of 
the field direction is resolved into more than 1000 time steps. 
An explicit Euler update scheme is used for integration and the 
simulations run until the crossover to asymptotic diffusion 
occurs or the total number of time steps exceeds $5{\times}10^5$. 

The structural disorder is effectively introduced by random 
displacements of magnetic poles from their ordered lattice positions.
The strength of disorder can be quantified by the parameter $\delta$, 
denoting the maximum possible displacement of each magnetic pole from 
its ordered lattice position. We randomly choose the position of 
each node within a circle of radius $L{\cdot} \delta$ around the 
corresponding node of the ordered triangular lattice, where $L$ 
denotes the size of the Wigner-Seitz cell. The dynamic disorder 
is generated by instantaneous random rearrangements of the poles 
at each switching event. Thus, the poles move to new positions 
where they stay immobile until the next switching event. The new 
random position of each pole remains within the allowed range 
around the original position of the corresponding ordered lattice 
node. This way we keep the deviations dynamic but smaller than 
$\delta{\cdot}L$ in all cases.

\begin{figure}[t]
\begin{center}
\includegraphics[width=\columnwidth,keepaspectratio]{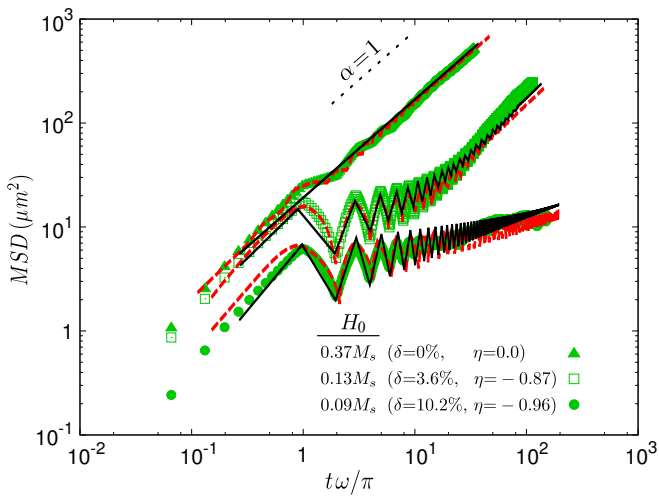}
\caption{MSD versus time, obtained from experiments (green symbols),
simulations (red dashed lines), and the persistent random walk model via
Eq.~(\ref{mean-r2}) (black solid lines). The experimental data at $\omega
{=} 15.7 \, {\rm rad \, s^{-1}}$ and various amplitudes $H_0$ are compared
to the best fits obtained by tunning a single fitting parameter (i.e.\
the lattice disorder $\delta$ in simulations, or the turning-angle
anisotropy $\eta$ in the analytical model). The best-fitting parameters
for each value of $H_0$ are given.}
\label{Fig:5}
\end{center}
\end{figure}

\begin{figure}[t]
\begin{center}
\includegraphics[width=\columnwidth,keepaspectratio]{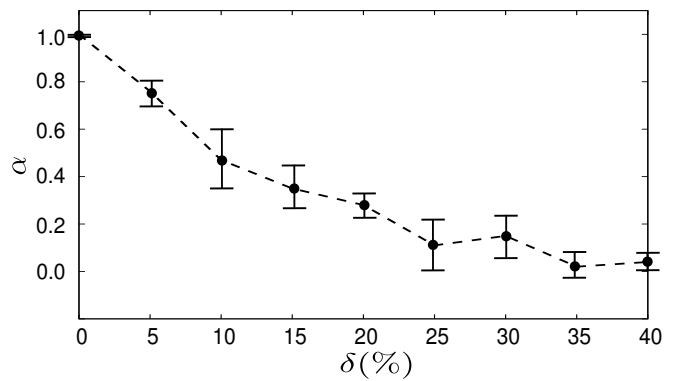}
\caption{The anomalous exponent $\alpha$ at short times (after smoothening
the MSD curves) in terms of the structural disorder $\delta$ of the lattice.}
\label{Fig:6}
\end{center}
\end{figure}

We analyze the particle trajectory for different values of $\delta$.
As shown in Fig.~\ref{Fig:4}, with increasing $\delta$ the 
distortion of the lattice is more pronounced, thus, the particle 
experiences oscillatory movements in traps more frequently and 
for longer times, which increases the asymmetry of the turning-angle 
distribution $f(\phi)$. By tunning $\delta$, as a single free 
parameter in our simulations, a remarkable agreement with the 
experimental MSD data is achieved (see Fig.~\ref{Fig:5}). 
Moreover, by smoothening the MSD curves obtained from simulations, 
we fit the initial slope of the curves to $\text{MSD}(t) {\sim} 
t^{\alpha}$ to get the anomalous exponent at short times. The 
results shown in Fig.~\ref{Fig:6} reveal that with increasing 
the amount of structural disorder, the slope gradually decreases 
from $\alpha{=}1$ for normal diffusion towards $\alpha{=}0$ 
corresponding to pure localization. The growth of MSD is extremely 
slow for $\delta{\geq}25\%$, in accord with the experimental 
data at weak fields $H_0 {<} 0.09 \, {\rm M_s}$.

\section{Persistent random walk model}
\label{Sec:PRW}
So far, we have shown that for small amplitudes of the external field
the turning-angle distribution $f(\phi)$ is asymmetric, which is accompanied 
by a subdiffusive dynamics at short times. In this section, we theoretically 
consider the particle motion at the level of individual steps and show
how the asymmetry of $f(\phi)$ with a tendency towards backward directions 
leads to subdiffusion. It was demonstrated in a previous work \cite{Tie09}
that the statistical properties of the subdiffusive motion could be captured 
by the ``{\it random walk on random walk}" model \cite{Keh82} (RWRW), which 
is a simple example of stochastic motion in a complex environment. In the 
RWRW model, the walker performs an ordinary random walk on an environment 
which is constructed by an ordinary random walk process as well. While this 
model showed a quantitative agreement with the experiments for certain 
field parameters \cite{Tie09}, the randomness of the environment indeed 
varies depending on the choice of $H_0$. Thus a more general theoretical 
framework is needed in order to capture the particle dynamics in the 
subdiffusive regime for all field parameters. Here, we look at the 
individual steps of the random walker and present an antipersistent model 
which enables us to reproduce fine details of motion such as the oscillatory 
dynamics observed at smaller values of $H_0$. Persistent random walk 
models \cite{Fur17,Tay22,Hau87,Wei94} have been used to describe e.g.\ 
stochastic transport on cytoskeletal filaments \cite{Sha14,Sad15}, 
self-propulsion subject to fluctuations \cite{How07,Per07}, or diffusive 
transport of light in foams and granular media \cite{Sad08,Sad11}. In 
the following, the asymmetry of $f(\phi)$ is quantified with $\eta 
{\equiv} \langle \cos\phi \rangle$, which varies from $0$ for an isotropic 
distribution $f(\phi){=}\frac{1}{2\pi}$ to $-1$ for an extremely asymmetric 
distribution $f(\phi){=}\delta(\phi{-}\pi)$. We obtain analytical 
expressions for the time evolution of the MSD, the crossover time to 
asymptotic diffusion, and the long-term diffusion coefficient in terms 
of the control parameter $\eta$ and compare the analytical predictions 
with the experimental data.

\begin{figure}[t]
\begin{center}
\includegraphics[width=\columnwidth,keepaspectratio]{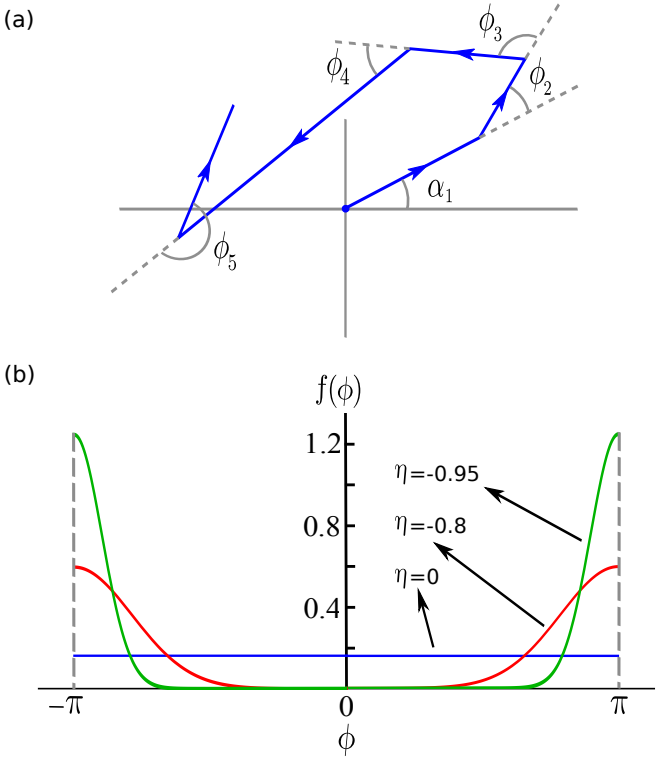}
\caption{(a) Trajectory of the walker during a few successive steps. (b)
Examples of the turning-angle distribution $f(\phi)$ for different 
values of the asymmetry measure $\eta$.}
\label{Fig:7}
\end{center}
\end{figure}

We consider a random walk in 2D, with uncorrelated step sizes $\ell$ which 
are obtained from an arbitrary distribution $g(\ell)$. The successive step
orientations are however correlated such that a new orientation is obtained 
from a turning-angle distribution $f(\phi)$ along the previous direction of 
motion (see Fig.~\ref{Fig:7}). While an isotropic $f(\phi)$ leads to a 
normal diffusion, introducing anisotropy along the arrival direction induces 
persistency and results in an anomalous transport on short time scales. The 
choices of $f(\phi)$ which encourage the walker towards backward (forward) 
directions lead to subdiffusion (superdiffusion). Such an approach had been 
used e.g.\ in the context of cell migration along surfaces \cite{Nos74}, 
animal movement \cite{Kareiva83}, and dynamics of polymer chains \cite{Flory69,Freed87}. 
Starting from the origin, let us assume that the first direction of 
motion is chosen randomly, i.e.\ the initial condition is $P(\alpha_1){=}
\frac{1}{2\pi}$. The $x$-coordinate of walker after $n$ steps can be 
obtained by projecting each step along the $x$-axis 
\begin{equation}
x = \sum_{i{=}1}^n x_i = \sum_{i{=}1}^n \ell_i \cos\alpha_i,
\label{x-projection}
\end{equation}
where $\alpha_i{=}\alpha_1{+}\phi_2{+}\cdot\cdot\cdot{+}\phi_i$. For simplicity, 
in the following we consider a constant step length $g(\ell){=}\delta(\ell{-}
\langle \ell \rangle)$ and $f(\phi)$ distributions with left-right symmetry. 
The first moment of displacement $\langle x \rangle$ then reads
\begin{equation}
\begin{aligned}
\langle x \rangle = &\langle \ell \rangle \langle \sum_{i{=}1}^n \cos\alpha_i 
\rangle =\langle \ell \rangle  \sum_{i{=}1}^n \int_{-\pi}^{\pi} \!\!\!\!\!\! 
d\phi_i f(\phi_i) \cdot\cdot\cdot \!\! \int_{-\pi}^{\pi} \!\!\!\!\!\! d\phi_2 
f(\phi_2) \\
&\times \int_{-\pi}^{\pi} \!\! d\alpha_1 P(\alpha_1) \cos(\alpha_1{+}\phi_2{+} 
\cdot\cdot\cdot\phi_i) = 0,
\label{mean-x}
\end{aligned}
\end{equation}
since the integral over $\alpha_1$ vanishes. Here $\langle \cdot\cdot\cdot 
\rangle$ denotes an ensemble average. The second moment of displacement is 
given by \cite{Nos74,Kareiva83,Flory69,Freed87}
\begin{equation}
\begin{aligned}
\langle x^2 \rangle &= \langle \sum_{i{=}1}^n \sum_{j{=}1}^n x_i x_j \rangle 
= \langle \sum_{i{=}1}^n x_i^2 +2 \sum_{i{>}j} x_i x_j \rangle \\
&= \sum_{i{=}1}^n \langle x_i^2 \rangle + 2 \sum_{i{=}1}^n \sum_{j{=}1}^{i{-}1} 
\langle x_i x_j \rangle.
\label{mean-x2-1}
\end{aligned}
\end{equation}
Similar to Eq.~(\ref{mean-x}), the first term can be obtained as 
\begin{figure}[t]
\begin{center}
\includegraphics[width=\columnwidth,keepaspectratio]{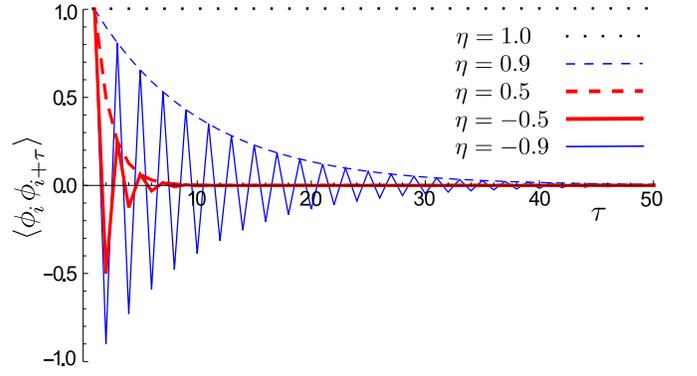}
\caption{The normalized auto-correlation function $\langle \phi_i \phi_{i{+}\tau}
\rangle$ versus time for different values of $\eta$. Positive (negative) $\eta$
corresponds to persistent (antipersistent) motion. $\eta{=}1$ denotes a 
ballistic motion without any turning.}
\label{Fig:8}
\end{center}
\end{figure}
\begin{equation}
\begin{aligned}
\langle x_i^2 \rangle = &\langle \ell^2 \rangle \int_{-\pi}^{\pi} \!\!\!\!\!\! 
d\phi_i f(\phi_i) \cdot\cdot\cdot \!\! \int_{-\pi}^{\pi} \!\!\!\!\!\! d\phi_2 
f(\phi_2) \times \\
&\int_{-\pi}^{\pi} \!\!\! d\alpha_1 P(\alpha_1) \cos^{2}(\alpha_1{+}\phi_2 
{+}\cdot\cdot\cdot\phi_i) = \frac{\langle \ell^2 \rangle}{2}.
\label{mean-x2-2}
\end{aligned}
\end{equation}
the second term on the right hand side of Eq.~(\ref{mean-x2-1}) can be 
evaluated in the following way
\begin{equation}
\begin{aligned}
&\langle x_i x_j \rangle = \langle \ell \rangle^2 \int_{-\pi}^{\pi} \!\!\!\!\!\! 
d\phi_i f(\phi_i) \cdot\cdot\cdot \!\! \int_{-\pi}^{\pi} \!\!\!\!\!\! d\phi_2 
f(\phi_2) \times \\
&\int_{-\pi}^{\pi} \!\!\! d\alpha_1 P(\alpha_1) \cos(\alpha_1 {+} \phi_2 {+} 
\cdot\cdot\cdot\phi_i) \cos(\alpha_1{+}\phi_2{+}\cdot\cdot\cdot\phi_j) \\
&=\frac{\langle \ell \rangle^2}{2} \int_{-\pi}^{\pi} \!\!\!\!\!\! d\phi_i 
f(\phi_i) \cdot\cdot\cdot d\phi_2 f(\phi_2) \cos(\phi_{j{+}1}{+}\cdot\cdot\cdot
\phi_i) \\
&=\frac{\langle \ell \rangle^2}{2} \Bigg( \int_{-\pi}^{\pi} \!\!\!\!\!\! d\phi 
f(\phi) \text{Re}[e^{i\phi}]\Bigg)^{i{-}j} = \frac{\langle \ell \rangle^2}{2} 
\eta^{i{-}j},
\label{mean-x2-3}
\end{aligned}
\end{equation}
where $\eta{=}\int_{-\pi}^{\pi} \!\! d\phi f(\phi) \cos\phi = \langle \cos\phi 
\rangle$. A few examples of the turning-angle distribution $f(\phi)$ and the
corresponding values of the asymmetry parameter $\eta$ are shown in 
Fig.~\ref{Fig:7}(b). From Eqs.~(\ref{mean-x2-1}) to (\ref{mean-x2-3}) one can
obtain the second moment $\langle x^2 \rangle$. Similar conclusions for the 
moments of the $y$-coordinate can be drawn due to symmetry. Thus, one gets 
the following expression for the mean squared displacement  
\begin{equation}
\begin{aligned}
\langle r^2 \rangle = \Big( \langle \ell^2 \rangle {+} \langle 
\ell \rangle^2 \frac{2 \eta}{1{-}\eta} \Big) n + \langle \ell \rangle^2 
\frac{2 \eta}{(1{-}\eta)^2} (\eta^n{-}1).
\label{mean-r2}
\end{aligned}
\end{equation}
The case $\eta{=}0$ corresponds to an isotropic distribution $f(\phi)$, for 
which Eq.~(\ref{mean-r2}) reduces to $\langle r^2 \rangle {=} \langle \ell^2 
\rangle n$, i.e.\ a normal diffusion. Negative (positive) values of $\eta$ denote 
an increased probability for motion in the near backward (forward) directions, 
thus, leading to antipersistent (persistent) motion. For $\eta {>} 0$ one obtains 
a superdiffusive short-time dynamics, while $\eta {<} 0$ leads to subdiffusion 
or oscillatory dynamics. In Fig.~\ref{Fig:5}, the theoretical predictions for
the MSD via Eq.~(\ref{mean-r2}) are compared with the experimental and simulation 
results. It can be seen that our theoretical approach remarkably reproduces 
the observed behavior by fitting the single free parameter of the model, i.e.\ 
$\eta$. Notably, the overall behavior of the MSD, the frequency of oscillations, 
the crossover time, and even the asymptotic diffusion coefficient are all 
captured by the theory.

The persistent random walker has a finite-range memory, beyond which the 
direction of motion gets completely randomized. According to Eq.~(\ref{mean-x2-3}), 
the auto-correlation between step orientations is given by
\begin{equation}
\langle \phi_i \phi_{i{+}\tau} \rangle = \langle \cos\phi \rangle^\tau,
\label{auto-correlation}
\end{equation}
which vanishes in the limit of $\tau {\rightarrow} \infty$ since $|\langle \cos\phi 
\rangle| {\leq} 1$. Some examples are shown in Fig.~\ref{Fig:8} for various values 
of $\eta$. One can estimate the crossover time $n_c$ to asymptotic diffusion 
by numerically solving $\Big( \langle \ell^2 \rangle {+} \langle \ell \rangle^2 
\frac{2 \eta}{1{-}\eta} \Big) n_c {\sim} \langle \ell \rangle^2 \frac{2 \eta}{(1
{-}\eta)^2} \eta^{n_c}.$ Moreover, it can be seen from Eq.~(\ref{mean-r2}) that 
the asymptotic diffusion coefficient also depends on $\eta$ as 
\begin{equation}
D_{\text{asymp}} = \frac14 v \Big( \frac{\langle \ell^2 \rangle}{\langle \ell 
\rangle} {+} \langle \ell \rangle \frac{2 \eta}{1{-}\eta} \Big),
\label{Dasymp}
\end{equation}
with $v$ being the average particle velocity. For a constant step size,
$D_{\text{asymp}}$ ranges from $0$ for $\eta{=}-1$ to $\frac14 v \langle 
\ell \rangle$ for $\eta{=}0$.

\section{Conclusions}
\label{Sec:Conclusions}
We combined experiment and theory to investigate the dynamics of paramagnetic
colloids driven above a two-state flashing potential. This potential was 
realized by periodically modulating the stray field generated at the surface 
of a magnetic bubble lattice in an uniaxial garnet film. The particles 
experience either enhanced diffusive or anomalous sub-diffusive dynamics, 
depending on the strength of the external drive. Applying a strong field 
leads to an ordinary random walk on the magnetic bubble lattice, while weaker 
fields result in structural disorder in the lattice which slows the particle 
dynamics. By means of a persistent random walk approach and numerical 
simulations we verified that increasing lattice disorder sharpens the 
turning-angle distribution of the particle towards backward directions, 
which decreases the anomalous exponent, postpones the crossover time to 
asymptotic diffusion, and modifies the long-term diffusion coefficient.

\bigskip
\noindent{{\bf Acknowledgment}\\
P. T. acknowledges support from the European Research Council Project No.\ 
335040, from Mineco (Grant No.\ FIS2013-41144-P and RYC-2011-07605), and 
AGAUR (Grant No.\ 2014SGR878). M.R.S.\ acknowledges support by the Deutsche 
Forschungsgemeinschaft (DFG) through Collaborative Research Centers SFB 
1027 (A7).}

\footnotesize{
\bibliography{Biblio} 

\begin{thebibliography}{46}
\expandafter\ifx\csname natexlab\endcsname\relax\def\natexlab#1{#1}\fi
\expandafter\ifx\csname bibnamefont\endcsname\relax
  \def\bibnamefont#1{#1}\fi
\expandafter\ifx\csname bibfnamefont\endcsname\relax
  \def\bibfnamefont#1{#1}\fi
\expandafter\ifx\csname citenamefont\endcsname\relax
  \def\citenamefont#1{#1}\fi
\expandafter\ifx\csname url\endcsname\relax
  \def\url#1{\texttt{#1}}\fi
\expandafter\ifx\csname urlprefix\endcsname\relax\def\urlprefix{URL }\fi
\providecommand{\bibinfo}[2]{#2}
\providecommand{\eprint}[2][]{\url{#2}}

\bibitem[{\citenamefont{Reimann}(2002)}]{Reim02}
\bibinfo{author}{\bibfnamefont{P.}~\bibnamefont{Reimann}},
  \bibinfo{journal}{Phys. Rep.} \textbf{\bibinfo{volume}{361}},
  \bibinfo{pages}{57} (\bibinfo{year}{2002}).

\bibitem[{\citenamefont{Hanggi and Marchesoni}(2009)}]{Han09}
\bibinfo{author}{\bibfnamefont{P.}~\bibnamefont{Hanggi}} \bibnamefont{and}
  \bibinfo{author}{\bibfnamefont{F.}~\bibnamefont{Marchesoni}},
  \bibinfo{journal}{Rev. Mod. Phys.} \textbf{\bibinfo{volume}{81}},
  \bibinfo{pages}{387} (\bibinfo{year}{2009}).

\bibitem[{\citenamefont{Sancho et~al.}(2004)\citenamefont{Sancho, Lacasta,
  Lindenberg, Sokolov, and Romero}}]{San04}
\bibinfo{author}{\bibfnamefont{J.~M.} \bibnamefont{Sancho}},
  \bibinfo{author}{\bibfnamefont{A.~M.} \bibnamefont{Lacasta}},
  \bibinfo{author}{\bibfnamefont{K.}~\bibnamefont{Lindenberg}},
  \bibinfo{author}{\bibfnamefont{I.~M.} \bibnamefont{Sokolov}},
  \bibnamefont{and} \bibinfo{author}{\bibfnamefont{A.~H.}
  \bibnamefont{Romero}}, \bibinfo{journal}{Phys. Rev. Lett.}
  \textbf{\bibinfo{volume}{92}}, \bibinfo{pages}{250601}
  (\bibinfo{year}{2004}).

\bibitem[{\citenamefont{Hanes et~al.}(2012)\citenamefont{Hanes, Dalle-Ferrier,
  Schmiedeberg, Jenkins, and Egelhaaf}}]{Han12}
\bibinfo{author}{\bibfnamefont{R.~D.~L.} \bibnamefont{Hanes}},
  \bibinfo{author}{\bibfnamefont{C.}~\bibnamefont{Dalle-Ferrier}},
  \bibinfo{author}{\bibfnamefont{M.}~\bibnamefont{Schmiedeberg}},
  \bibinfo{author}{\bibfnamefont{M.~C.} \bibnamefont{Jenkins}},
  \bibnamefont{and} \bibinfo{author}{\bibfnamefont{S.~U.}
  \bibnamefont{Egelhaaf}}, \bibinfo{journal}{Soft Matter}
  \textbf{\bibinfo{volume}{109}}, \bibinfo{pages}{10245}
  (\bibinfo{year}{2012}).

\bibitem[{\citenamefont{Villegas et~al.}(2003)\citenamefont{Villegas,
  Savel\'{e}v, Nori, Gonzalez, Anguita, Garcia, and Vicent}}]{Vil03}
\bibinfo{author}{\bibfnamefont{J.~E.} \bibnamefont{Villegas}},
  \bibinfo{author}{\bibfnamefont{S.}~\bibnamefont{Savel\'{e}v}},
  \bibinfo{author}{\bibfnamefont{F.}~\bibnamefont{Nori}},
  \bibinfo{author}{\bibfnamefont{E.~M.} \bibnamefont{Gonzalez}},
  \bibinfo{author}{\bibfnamefont{J.~V.} \bibnamefont{Anguita}},
  \bibinfo{author}{\bibfnamefont{R.}~\bibnamefont{Garcia}}, \bibnamefont{and}
  \bibinfo{author}{\bibfnamefont{J.~L.} \bibnamefont{Vicent}},
  \bibinfo{journal}{Science} \textbf{\bibinfo{volume}{302}},
  \bibinfo{pages}{1188} (\bibinfo{year}{2003}).

\bibitem[{\citenamefont{de~Vondel et~al.}(2005)\citenamefont{de~Vondel,
  de~Souza~Silva, Zhu, Morelle, and Moshchalkov}}]{Von05}
\bibinfo{author}{\bibfnamefont{J.~V.} \bibnamefont{de~Vondel}},
  \bibinfo{author}{\bibfnamefont{C.~C.} \bibnamefont{de~Souza~Silva}},
  \bibinfo{author}{\bibfnamefont{B.~Y.} \bibnamefont{Zhu}},
  \bibinfo{author}{\bibfnamefont{M.}~\bibnamefont{Morelle}}, \bibnamefont{and}
  \bibinfo{author}{\bibfnamefont{V.~V.} \bibnamefont{Moshchalkov}},
  \bibinfo{journal}{Phys. Rev. Lett.} \textbf{\bibinfo{volume}{94}},
  \bibinfo{pages}{057003} (\bibinfo{year}{2005}).

\bibitem[{\citenamefont{Majer et~al.}(2003)\citenamefont{Majer, Peguiron,
  Grifoni, Tusveld, and Mooij}}]{Maj03}
\bibinfo{author}{\bibfnamefont{J.~B.} \bibnamefont{Majer}},
  \bibinfo{author}{\bibfnamefont{J.}~\bibnamefont{Peguiron}},
  \bibinfo{author}{\bibfnamefont{M.}~\bibnamefont{Grifoni}},
  \bibinfo{author}{\bibfnamefont{M.}~\bibnamefont{Tusveld}}, \bibnamefont{and}
  \bibinfo{author}{\bibfnamefont{J.~E.} \bibnamefont{Mooij}},
  \bibinfo{journal}{Phys. Rev. Lett.} \textbf{\bibinfo{volume}{90}},
  \bibinfo{pages}{056802} (\bibinfo{year}{2003}).

\bibitem[{\citenamefont{Ustinov et~al.}(2004)\citenamefont{Ustinov, Coqui,
  Kemp, Zolotaryuk, and Salerno}}]{Ust04}
\bibinfo{author}{\bibfnamefont{A.~V.} \bibnamefont{Ustinov}},
  \bibinfo{author}{\bibfnamefont{C.}~\bibnamefont{Coqui}},
  \bibinfo{author}{\bibfnamefont{A.}~\bibnamefont{Kemp}},
  \bibinfo{author}{\bibfnamefont{Y.}~\bibnamefont{Zolotaryuk}},
  \bibnamefont{and} \bibinfo{author}{\bibfnamefont{M.}~\bibnamefont{Salerno}},
  \bibinfo{journal}{Phys. Rev. Lett.} \textbf{\bibinfo{volume}{93}},
  \bibinfo{pages}{087001} (\bibinfo{year}{2004}).

\bibitem[{\citenamefont{Mahmud et~al.}(2009)\citenamefont{Mahmud, Campbell,
  Bishop, Komarova, Chaga, Soh, Huda, Kandere-Grzybowska, and
  Grzybowski}}]{Mah09}
\bibinfo{author}{\bibfnamefont{G.}~\bibnamefont{Mahmud}},
  \bibinfo{author}{\bibfnamefont{C.~J.} \bibnamefont{Campbell}},
  \bibinfo{author}{\bibfnamefont{K.~J.~M.} \bibnamefont{Bishop}},
  \bibinfo{author}{\bibfnamefont{Y.~A.} \bibnamefont{Komarova}},
  \bibinfo{author}{\bibfnamefont{O.}~\bibnamefont{Chaga}},
  \bibinfo{author}{\bibfnamefont{S.}~\bibnamefont{Soh}},
  \bibinfo{author}{\bibfnamefont{S.}~\bibnamefont{Huda}},
  \bibinfo{author}{\bibfnamefont{K.}~\bibnamefont{Kandere-Grzybowska}},
  \bibnamefont{and} \bibinfo{author}{\bibfnamefont{B.~A.}
  \bibnamefont{Grzybowski}}, \bibinfo{journal}{Nat. Phys.}
  \textbf{\bibinfo{volume}{5}}, \bibinfo{pages}{606} (\bibinfo{year}{2009}).

\bibitem[{\citenamefont{Julicher et~al.}(1997)\citenamefont{Julicher, Ajdari,
  and Prost}}]{Jul97}
\bibinfo{author}{\bibfnamefont{F.}~\bibnamefont{Julicher}},
  \bibinfo{author}{\bibfnamefont{A.}~\bibnamefont{Ajdari}}, \bibnamefont{and}
  \bibinfo{author}{\bibfnamefont{J.}~\bibnamefont{Prost}},
  \bibinfo{journal}{Rev. Mod. Phys.} \textbf{\bibinfo{volume}{69}},
  \bibinfo{pages}{1269} (\bibinfo{year}{1997}).

\bibitem[{\citenamefont{Korda et~al.}(2002)\citenamefont{Korda, Taylor, and
  Grier}}]{Kor02}
\bibinfo{author}{\bibfnamefont{P.~T.} \bibnamefont{Korda}},
  \bibinfo{author}{\bibfnamefont{M.~B.} \bibnamefont{Taylor}},
  \bibnamefont{and} \bibinfo{author}{\bibfnamefont{D.~G.} \bibnamefont{Grier}},
  \bibinfo{journal}{Phys. Rev. Lett.} \textbf{\bibinfo{volume}{89}},
  \bibinfo{pages}{128301} (\bibinfo{year}{2002}).

\bibitem[{\citenamefont{MacDonald et~al.}(2003)\citenamefont{MacDonald,
  Spalding, and Dholakia}}]{Mac03}
\bibinfo{author}{\bibfnamefont{M.~P.} \bibnamefont{MacDonald}},
  \bibinfo{author}{\bibfnamefont{G.~C.} \bibnamefont{Spalding}},
  \bibnamefont{and} \bibinfo{author}{\bibfnamefont{K.}~\bibnamefont{Dholakia}},
  \bibinfo{journal}{Nature} \textbf{\bibinfo{volume}{426}},
  \bibinfo{pages}{421} (\bibinfo{year}{2003}).

\bibitem[{\citenamefont{Huang et~al.}(2004)\citenamefont{Huang, Cox, Austin,
  and Sturm}}]{Hua04}
\bibinfo{author}{\bibfnamefont{L.~R.} \bibnamefont{Huang}},
  \bibinfo{author}{\bibfnamefont{E.~C.} \bibnamefont{Cox}},
  \bibinfo{author}{\bibfnamefont{R.~H.} \bibnamefont{Austin}},
  \bibnamefont{and} \bibinfo{author}{\bibfnamefont{J.~C.} \bibnamefont{Sturm}},
  \bibinfo{journal}{Science} \textbf{\bibinfo{volume}{304}},
  \bibinfo{pages}{987} (\bibinfo{year}{2004}).

\bibitem[{\citenamefont{Lacasta et~al.}(2005)\citenamefont{Lacasta, Sancho,
  Romero, and Lindenberg}}]{Lac05}
\bibinfo{author}{\bibfnamefont{A.~M.} \bibnamefont{Lacasta}},
  \bibinfo{author}{\bibfnamefont{J.~M.} \bibnamefont{Sancho}},
  \bibinfo{author}{\bibfnamefont{A.~H.} \bibnamefont{Romero}},
  \bibnamefont{and}
  \bibinfo{author}{\bibfnamefont{K.}~\bibnamefont{Lindenberg}},
  \bibinfo{journal}{Phys. Rev. Lett.} \textbf{\bibinfo{volume}{94}},
  \bibinfo{pages}{160601} (\bibinfo{year}{2005}).

\bibitem[{\citenamefont{Tierno et~al.}(2008)\citenamefont{Tierno, Soba,
  Johansen, and Sagu\'{e}s}}]{Tie08}
\bibinfo{author}{\bibfnamefont{P.}~\bibnamefont{Tierno}},
  \bibinfo{author}{\bibfnamefont{A.}~\bibnamefont{Soba}},
  \bibinfo{author}{\bibfnamefont{T.~H.} \bibnamefont{Johansen}},
  \bibnamefont{and}
  \bibinfo{author}{\bibfnamefont{F.}~\bibnamefont{Sagu\'{e}s}},
  \bibinfo{journal}{Appl. Phys. Lett.} \textbf{\bibinfo{volume}{93}},
  \bibinfo{pages}{214102} (\bibinfo{year}{2008}).

\bibitem[{\citenamefont{Dalle-Ferrier et~al.}(2011)\citenamefont{Dalle-Ferrier,
  Kruger, Hanes, Walta, Jenkins, and Egelhaaf}}]{Dal11}
\bibinfo{author}{\bibfnamefont{C.}~\bibnamefont{Dalle-Ferrier}},
  \bibinfo{author}{\bibfnamefont{M.}~\bibnamefont{Kruger}},
  \bibinfo{author}{\bibfnamefont{R.~D.~L.} \bibnamefont{Hanes}},
  \bibinfo{author}{\bibfnamefont{S.}~\bibnamefont{Walta}},
  \bibinfo{author}{\bibfnamefont{M.~C.} \bibnamefont{Jenkins}},
  \bibnamefont{and} \bibinfo{author}{\bibfnamefont{S.~U.}
  \bibnamefont{Egelhaaf}}, \bibinfo{journal}{Soft Matter}
  \textbf{\bibinfo{volume}{7}}, \bibinfo{pages}{2064} (\bibinfo{year}{2011}).

\bibitem[{\citenamefont{Tierno et~al.}(2007)\citenamefont{Tierno, Johansen, and
  Fischer}}]{Tie07}
\bibinfo{author}{\bibfnamefont{P.}~\bibnamefont{Tierno}},
  \bibinfo{author}{\bibfnamefont{T.~H.} \bibnamefont{Johansen}},
  \bibnamefont{and} \bibinfo{author}{\bibfnamefont{T.~M.}
  \bibnamefont{Fischer}}, \bibinfo{journal}{Phys. Rev. Lett.}
  \textbf{\bibinfo{volume}{99}}, \bibinfo{pages}{038303}
  (\bibinfo{year}{2007}).

\bibitem[{\citenamefont{Pertsinidis and Ling}(2008)}]{Per08}
\bibinfo{author}{\bibfnamefont{A.}~\bibnamefont{Pertsinidis}} \bibnamefont{and}
  \bibinfo{author}{\bibfnamefont{X.~S.} \bibnamefont{Ling}},
  \bibinfo{journal}{Phys. Rev. Lett.} \textbf{\bibinfo{volume}{100}},
  \bibinfo{pages}{028303} (\bibinfo{year}{2008}).

\bibitem[{\citenamefont{Astumian and Bier}(1994)}]{Ast94}
\bibinfo{author}{\bibfnamefont{R.~D.} \bibnamefont{Astumian}} \bibnamefont{and}
  \bibinfo{author}{\bibfnamefont{M.}~\bibnamefont{Bier}},
  \bibinfo{journal}{Phys. Rev. Lett.} \textbf{\bibinfo{volume}{72}},
  \bibinfo{pages}{1766} (\bibinfo{year}{1994}).

\bibitem[{\citenamefont{Prost et~al.}(1994)\citenamefont{Prost, Chauwin,
  Peliti, and Adjari}}]{Pro94}
\bibinfo{author}{\bibfnamefont{J.}~\bibnamefont{Prost}},
  \bibinfo{author}{\bibfnamefont{J.~F.} \bibnamefont{Chauwin}},
  \bibinfo{author}{\bibfnamefont{L.}~\bibnamefont{Peliti}}, \bibnamefont{and}
  \bibinfo{author}{\bibfnamefont{A.}~\bibnamefont{Adjari}},
  \bibinfo{journal}{Phys. Rev. Lett.} \textbf{\bibinfo{volume}{72}},
  \bibinfo{pages}{2652} (\bibinfo{year}{1994}).

\bibitem[{\citenamefont{Cao et~al.}(2004)\citenamefont{Cao, Dinis, and
  Parrondo}}]{Cao04}
\bibinfo{author}{\bibfnamefont{F.~J.} \bibnamefont{Cao}},
  \bibinfo{author}{\bibfnamefont{L.}~\bibnamefont{Dinis}}, \bibnamefont{and}
  \bibinfo{author}{\bibfnamefont{J.~M.~R.} \bibnamefont{Parrondo}},
  \bibinfo{journal}{Phys. Rev. Lett.} \textbf{\bibinfo{volume}{93}},
  \bibinfo{pages}{040603} (\bibinfo{year}{2004}).

\bibitem[{\citenamefont{Faucheux et~al.}(1995)\citenamefont{Faucheux, Bourdieu,
  Kaplan, and Libchaber}}]{Fau95}
\bibinfo{author}{\bibfnamefont{L.~P.} \bibnamefont{Faucheux}},
  \bibinfo{author}{\bibfnamefont{L.~S.} \bibnamefont{Bourdieu}},
  \bibinfo{author}{\bibfnamefont{P.~D.} \bibnamefont{Kaplan}},
  \bibnamefont{and} \bibinfo{author}{\bibfnamefont{A.~J.}
  \bibnamefont{Libchaber}}, \bibinfo{journal}{Phys. Rev. Lett.}
  \textbf{\bibinfo{volume}{74}}, \bibinfo{pages}{1504} (\bibinfo{year}{1995}).

\bibitem[{\citenamefont{Lib\'{a}l et~al.}(2006)\citenamefont{Lib\'{a}l,
  Reichhardt, Jank\'{o}, and Reichhardt}}]{Lib06}
\bibinfo{author}{\bibfnamefont{A.}~\bibnamefont{Lib\'{a}l}},
  \bibinfo{author}{\bibfnamefont{C.}~\bibnamefont{Reichhardt}},
  \bibinfo{author}{\bibfnamefont{B.}~\bibnamefont{Jank\'{o}}},
  \bibnamefont{and} \bibinfo{author}{\bibfnamefont{C.~J.~O.}
  \bibnamefont{Reichhardt}}, \bibinfo{journal}{Phys. Rev. Lett.}
  \textbf{\bibinfo{volume}{96}}, \bibinfo{pages}{188301}
  (\bibinfo{year}{2006}).

\bibitem[{\citenamefont{Tierno et~al.}(2009)\citenamefont{Tierno, Sagu\'{e}s,
  Johansen, and Fischer}}]{Tie09}
\bibinfo{author}{\bibfnamefont{P.}~\bibnamefont{Tierno}},
  \bibinfo{author}{\bibfnamefont{F.}~\bibnamefont{Sagu\'{e}s}},
  \bibinfo{author}{\bibfnamefont{T.~H.} \bibnamefont{Johansen}},
  \bibnamefont{and} \bibinfo{author}{\bibfnamefont{T.~M.}
  \bibnamefont{Fischer}}, \bibinfo{journal}{Phys. Chem. Chem. Phys.}
  \textbf{\bibinfo{volume}{11}}, \bibinfo{pages}{9615} (\bibinfo{year}{2009}).

\bibitem[{\citenamefont{Eschenfelder}(1980)}]{Esc80}
\bibinfo{author}{\bibfnamefont{A.}~\bibnamefont{Eschenfelder}},
  \emph{\bibinfo{title}{Magnetic bubble technology}}
  (\bibinfo{publisher}{Springer-Verlag}, \bibinfo{address}{Berlin},
  \bibinfo{year}{1980}).

\bibitem[{\citenamefont{Crocker and Grier}(1996)}]{Cro96}
\bibinfo{author}{\bibfnamefont{J.~C.} \bibnamefont{Crocker}} \bibnamefont{and}
  \bibinfo{author}{\bibfnamefont{D.~G.} \bibnamefont{Grier}},
  \bibinfo{journal}{J. Colloid Interface Sci.} \textbf{\bibinfo{volume}{179}},
  \bibinfo{pages}{298} (\bibinfo{year}{1996}).

\bibitem[{\citenamefont{Soba et~al.}(2008)\citenamefont{Soba, Tierno, Fischer,
  and Sagu\'{e}s}}]{Sob08}
\bibinfo{author}{\bibfnamefont{A.}~\bibnamefont{Soba}},
  \bibinfo{author}{\bibfnamefont{P.}~\bibnamefont{Tierno}},
  \bibinfo{author}{\bibfnamefont{T.~M.} \bibnamefont{Fischer}},
  \bibnamefont{and}
  \bibinfo{author}{\bibfnamefont{F.}~\bibnamefont{Sagu\'{e}s}},
  \bibinfo{journal}{Phys. Rev. E} \textbf{\bibinfo{volume}{77}},
  \bibinfo{pages}{060401} (\bibinfo{year}{2008}).

\bibitem[{\citenamefont{Machta and Zwanzig}(1983)}]{Mac83}
\bibinfo{author}{\bibfnamefont{J.}~\bibnamefont{Machta}} \bibnamefont{and}
  \bibinfo{author}{\bibfnamefont{R.}~\bibnamefont{Zwanzig}},
  \bibinfo{journal}{Phys. Rev. Lett.} \textbf{\bibinfo{volume}{50}},
  \bibinfo{pages}{1959} (\bibinfo{year}{1983}).

\bibitem[{\citenamefont{Klages and Dellago}(2000)}]{Kla00}
\bibinfo{author}{\bibfnamefont{R.}~\bibnamefont{Klages}} \bibnamefont{and}
  \bibinfo{author}{\bibfnamefont{C.}~\bibnamefont{Dellago}},
  \bibinfo{journal}{J. Stat. Phys.} \textbf{\bibinfo{volume}{101}},
  \bibinfo{pages}{1145} (\bibinfo{year}{2000}).

\bibitem[{\citenamefont{Bar'yakhtar et~al.}(1977)\citenamefont{Bar'yakhtar,
  Gann, Gorobets, Smolenski, and Filippov}}]{Bar77}
\bibinfo{author}{\bibfnamefont{V.~G.} \bibnamefont{Bar'yakhtar}},
  \bibinfo{author}{\bibfnamefont{V.~V.} \bibnamefont{Gann}},
  \bibinfo{author}{\bibfnamefont{Y.~I.} \bibnamefont{Gorobets}},
  \bibinfo{author}{\bibfnamefont{G.~A.} \bibnamefont{Smolenski}},
  \bibnamefont{and} \bibinfo{author}{\bibfnamefont{B.~N.}
  \bibnamefont{Filippov}}, \bibinfo{journal}{Soviet Physics Uspekhi}
  \textbf{\bibinfo{volume}{20}}, \bibinfo{pages}{298} (\bibinfo{year}{1977}).

\bibitem[{\citenamefont{Seshadri and Westervelt}(1992)}]{Ses92}
\bibinfo{author}{\bibfnamefont{R.}~\bibnamefont{Seshadri}} \bibnamefont{and}
  \bibinfo{author}{\bibfnamefont{R.~M.} \bibnamefont{Westervelt}},
  \bibinfo{journal}{Phys. Rev. B} \textbf{\bibinfo{volume}{46}},
  \bibinfo{pages}{5142} (\bibinfo{year}{1992}).

\bibitem[{\citenamefont{Kehr and Kutner}(1982)}]{Keh82}
\bibinfo{author}{\bibfnamefont{K.~W.} \bibnamefont{Kehr}} \bibnamefont{and}
  \bibinfo{author}{\bibfnamefont{R.}~\bibnamefont{Kutner}},
  \bibinfo{journal}{Physica} \textbf{\bibinfo{volume}{110A}},
  \bibinfo{pages}{535} (\bibinfo{year}{1982}).

\bibitem[{\citenamefont{Furth}(1917)}]{Fur17}
\bibinfo{author}{\bibfnamefont{R.}~\bibnamefont{Furth}}, \bibinfo{journal}{Ann.
  Phys.} \textbf{\bibinfo{volume}{53}}, \bibinfo{pages}{177}
  (\bibinfo{year}{1917}).

\bibitem[{\citenamefont{Taylor}(1922)}]{Tay22}
\bibinfo{author}{\bibfnamefont{G.~I.} \bibnamefont{Taylor}},
  \bibinfo{journal}{Proceeding of the London Mathematical Society}
  \textbf{\bibinfo{volume}{20}}, \bibinfo{pages}{196} (\bibinfo{year}{1922}).

\bibitem[{\citenamefont{Haus and Kehr}(1987)}]{Hau87}
\bibinfo{author}{\bibfnamefont{J.~W.} \bibnamefont{Haus}} \bibnamefont{and}
  \bibinfo{author}{\bibfnamefont{K.~W.} \bibnamefont{Kehr}},
  \bibinfo{journal}{Phys. Rep.} \textbf{\bibinfo{volume}{150}},
  \bibinfo{pages}{263} (\bibinfo{year}{1987}).

\bibitem[{\citenamefont{Weiss}(1994)}]{Wei94}
\bibinfo{author}{\bibfnamefont{G.~H.} \bibnamefont{Weiss}},
  \emph{\bibinfo{title}{Aspects and Applications of the Random Walk}}
  (\bibinfo{publisher}{North-Holland}, \bibinfo{address}{Amsterdam},
  \bibinfo{year}{1994}).

\bibitem[{\citenamefont{Shaebani et~al.}(2014)\citenamefont{Shaebani, Sadjadi,
  Sokolov, Rieger, and Santen}}]{Sha14}
\bibinfo{author}{\bibfnamefont{M.~R.} \bibnamefont{Shaebani}},
  \bibinfo{author}{\bibfnamefont{Z.}~\bibnamefont{Sadjadi}},
  \bibinfo{author}{\bibfnamefont{I.~M.} \bibnamefont{Sokolov}},
  \bibinfo{author}{\bibfnamefont{H.}~\bibnamefont{Rieger}}, \bibnamefont{and}
  \bibinfo{author}{\bibfnamefont{L.}~\bibnamefont{Santen}},
  \bibinfo{journal}{Phys. Rev. E} \textbf{\bibinfo{volume}{90}},
  \bibinfo{pages}{030701(R)} (\bibinfo{year}{2014}).

\bibitem[{\citenamefont{Sadjadi et~al.}(2015)\citenamefont{Sadjadi, Shaebani,
  Rieger, and Santen}}]{Sad15}
\bibinfo{author}{\bibfnamefont{Z.}~\bibnamefont{Sadjadi}},
  \bibinfo{author}{\bibfnamefont{M.~R.} \bibnamefont{Shaebani}},
  \bibinfo{author}{\bibfnamefont{H.}~\bibnamefont{Rieger}}, \bibnamefont{and}
  \bibinfo{author}{\bibfnamefont{L.}~\bibnamefont{Santen}},
  \bibinfo{journal}{Phys. Rev. E} \textbf{\bibinfo{volume}{91}},
  \bibinfo{pages}{062715} (\bibinfo{year}{2015}).

\bibitem[{\citenamefont{Howse et~al.}(2007)\citenamefont{Howse, Jones, Ryan,
  Gough, Vafabakhsh, and Golestanian}}]{How07}
\bibinfo{author}{\bibfnamefont{J.~R.} \bibnamefont{Howse}},
  \bibinfo{author}{\bibfnamefont{R.~A.~L.} \bibnamefont{Jones}},
  \bibinfo{author}{\bibfnamefont{A.~J.} \bibnamefont{Ryan}},
  \bibinfo{author}{\bibfnamefont{T.}~\bibnamefont{Gough}},
  \bibinfo{author}{\bibfnamefont{R.}~\bibnamefont{Vafabakhsh}},
  \bibnamefont{and}
  \bibinfo{author}{\bibfnamefont{R.}~\bibnamefont{Golestanian}},
  \bibinfo{journal}{Phys. Rev. Lett.} \textbf{\bibinfo{volume}{99}},
  \bibinfo{pages}{048102} (\bibinfo{year}{2007}).

\bibitem[{\citenamefont{Peruani and Morelli}(2007)}]{Per07}
\bibinfo{author}{\bibfnamefont{F.}~\bibnamefont{Peruani}} \bibnamefont{and}
  \bibinfo{author}{\bibfnamefont{L.}~\bibnamefont{Morelli}},
  \bibinfo{journal}{Phys. Rev. Lett.} \textbf{\bibinfo{volume}{99}},
  \bibinfo{pages}{010602} (\bibinfo{year}{2007}).

\bibitem[{\citenamefont{Sadjadi et~al.}(2008)\citenamefont{Sadjadi, Miri,
  Shaebani, and Nakhaee}}]{Sad08}
\bibinfo{author}{\bibfnamefont{Z.}~\bibnamefont{Sadjadi}},
  \bibinfo{author}{\bibfnamefont{M.~F.} \bibnamefont{Miri}},
  \bibinfo{author}{\bibfnamefont{M.~R.} \bibnamefont{Shaebani}},
  \bibnamefont{and} \bibinfo{author}{\bibfnamefont{S.}~\bibnamefont{Nakhaee}},
  \bibinfo{journal}{Phys. Rev. E} \textbf{\bibinfo{volume}{78}},
  \bibinfo{pages}{031121} (\bibinfo{year}{2008}).

\bibitem[{\citenamefont{Sadjadi and Miri}(2011)}]{Sad11}
\bibinfo{author}{\bibfnamefont{Z.}~\bibnamefont{Sadjadi}} \bibnamefont{and}
  \bibinfo{author}{\bibfnamefont{M.~F.} \bibnamefont{Miri}},
  \bibinfo{journal}{Phys. Rev. E} \textbf{\bibinfo{volume}{84}},
  \bibinfo{pages}{051305} (\bibinfo{year}{2011}).

\bibitem[{\citenamefont{Nossal and Weiss}(1974)}]{Nos74}
\bibinfo{author}{\bibfnamefont{R.}~\bibnamefont{Nossal}} \bibnamefont{and}
  \bibinfo{author}{\bibfnamefont{G.~H.} \bibnamefont{Weiss}},
  \bibinfo{journal}{J. Theor. Biol.} \textbf{\bibinfo{volume}{47}},
  \bibinfo{pages}{103} (\bibinfo{year}{1974}).

\bibitem[{\citenamefont{Kareiva and Shigesada}(1983)}]{Kareiva83}
\bibinfo{author}{\bibfnamefont{P.~M.} \bibnamefont{Kareiva}} \bibnamefont{and}
  \bibinfo{author}{\bibfnamefont{N.}~\bibnamefont{Shigesada}},
  \bibinfo{journal}{Oecologia} \textbf{\bibinfo{volume}{56}},
  \bibinfo{pages}{234} (\bibinfo{year}{1983}).

\bibitem[{\citenamefont{Flory}(1969)}]{Flory69}
\bibinfo{author}{\bibfnamefont{P.~J.} \bibnamefont{Flory}},
  \emph{\bibinfo{title}{Statistical Mechanics of Chain Molecules}}
  (\bibinfo{publisher}{Wiley}, \bibinfo{address}{New York},
  \bibinfo{year}{1969}).

\bibitem[{\citenamefont{Freed}(1987)}]{Freed87}
\bibinfo{author}{\bibfnamefont{K.~F.} \bibnamefont{Freed}},
  \emph{\bibinfo{title}{Renormalization Group Theory of Macromolecules}}
  (\bibinfo{publisher}{Wiley}, \bibinfo{address}{New York},
  \bibinfo{year}{1987}).

\end{thebibliography}
\bibliographystyle{rsc}
}

\end{document}